\documentclass{article}

\usepackage{arxiv}

\usepackage[utf8]{inputenc} 
\usepackage[T1]{fontenc}    
\usepackage{hyperref}       
\usepackage{url}            
\usepackage{booktabs}       
\usepackage{amsfonts}       
\usepackage{nicefrac}       
\usepackage{microtype}      
\usepackage{lipsum}

\usepackage{graphicx}
\usepackage{amsmath,amssymb}
\pdfoutput=1
\usepackage{changepage}
\usepackage{textcomp,marvosym}
\usepackage[right]{lineno}
\usepackage{microtype}
\DisableLigatures[f]{encoding = *, family = * }
\usepackage[table]{xcolor}
\usepackage{array}

\newcolumntype{+}{!{\vrule width 2pt}}

\newlength\savedwidth

\makeatletter

\title{Short research review: Applications of statistical physics investigating financial and other social systems}

\author{
  Vygintas~Gontis, Aleksejus~Kononovičius, Julius~Ruseckas \\
  Institute of Theoretical Physics and Astronomy\\
  Vilnius University\\
  Saul{\. e}tekio al. 3, 10257 Vilnius \\
  \texttt{vygintas@gontis.eu} \\
}

\begin{document}
\maketitle

Physics research complements traditional approaches, such as mathematical (stochastic) finance and econometrics in quantitative economics and finance \cite{Mantegna2005QF,Chakrabarti2006Wiley}. In the early years of this millennium, we embarked on an interdisciplinary research endeavor in Lithuania, applying concepts from statistical physics to understand complex financial and social systems. Here, we provide a short review of investigations in Lithuania, spanning from 2008 to 2022, undertaken by our research group.

The term "Econophysics" was coined by Professor H. E. Stanley over three decades ago, merging the disciplines of Economics and Physics. While the precise definition of "Econophysics" remains subject to ongoing debate, numerous reputable research groups and scientific journals worldwide are actively contributing to its development.

Our journey into Econophysics prompts the intriguing question of how theories primarily designed to explain the physical world in terms of particles can be applied to unravel the complexities of human social and economic behavior. Physics, as a natural science, is renowned for its precision and predictive power, grounded in the utilization of a limited set of universal properties of matter, sufficient to elucidate a multitude of physical phenomena. In stark contrast, the social sciences lack equivalent precise universal properties for individuals. Human beings, unlike fundamental particles, differ significantly from one another in numerous aspects. Despite this fundamental distinction between physical and social systems, the analogy between statistical laws in physics and social sciences continues to drive the application of statistical physics to investigate the properties of financial and other social systems.

Statistical mechanics, a foundational branch of physics, enables the prediction and explanation of the measurable characteristics of macroscopic systems based on the behavior of their microscopic constituents. However, even in physical systems, the presence of dynamical chaos complicates the relationship between microscopic and macroscopic descriptions. In the realm of economic systems, it is nearly impossible to formulate microscopic equations of motion for all interacting entities. Therefore, researchers typically begin their investigations at the macro level. Consequently, understanding the global behavior of economic systems necessitates the application of concepts such as complex systems, stochastic dynamics, correlation effects, self-organization, self-similarity, and scaling. Remarkably, delving into detailed microscopic descriptions of economic systems is unnecessary for these applications. The concepts and methods of statistical physics, honed through their application to diverse and intricate systems, including financial systems, have proven invaluable. To elucidate empirical findings in social systems, physicists have developed sophisticated kinetic exchange models, offering intriguing parallels to established physical theories. Notably, the microscopic modeling of social systems, comprising intelligence-free agents exhibiting probabilistic behavior, has gained recognition as a viable investigative approach.

Our contributions to Econophysics encompass the application of statistical physics, empirical data analysis, stochastic modeling, agent-based modeling, and the synthesis of micro and macro approaches into a coherent framework. We initiate our research by leveraging extensive financial data and extend our inquiries to encompass various other social systems whenever data availability permits.

Our initial research efforts were dedicated to modeling financial markets \cite{Gontis2001LietFizRink,Gontis2002MC}. In these studies, we treated trades within financial markets as point events driven by a point process proposed in prior works \cite{Kaulakys1998PRE,Kaulakys1999PLA,Kaulakys2005PhysRevE}. While modeling trades as point events aligns with intuition, modeling volatility and return as point processes presented a more intricate challenge. To address this challenge, we refined our approach by abstracting from point processes and instead embraced a continuous framework founded on Langevin stochastic differential equations (SDEs) \cite{Ruseckas2010PhysRevE}. These nonlinear SDEs engender statistical properties in the variable, which can be interpreted as indicative of long-range memory. Consequently, the utilization of stochastic modeling rooted in nonlinear SDEs serves as evidence of the presence of spurious long-range memory \cite{Gontis2017Entropy,Gontis2018PhysA}.

Our journey led us to further enhance our nonlinear SDE approach by incorporating models for volatility and return \cite{Gontis2010Intech,Gontis2010PhysA,Gontis2011JDySES,Ruseckas2012ACS}. Interestingly, we discovered that similar SDEs could be derived from a straightforward agent-based model (ABM) \cite{Ruseckas2011EPL,Kononovicius2012PhysA}. Over time, our ABMs evolved in complexity to account for the separation of time scales and order flow \cite{Gontis2014PlosOne,Kononovicius2019OB}. We even ventured into the realm of sociophysics, recognizing the equivalence between our herding ABM used in financial market modeling and the well-known voter model \cite{Kononovicius2017Complexity,Kononovicius2019CompJStat,Kononovicius2020JStatMech,Kononovicius2021CSF}.

Conversations and collaborations with eminent figures in the field of Econophysics, including Professor H. E. Stanley, Professors Sh. Havlin, B. Podobnik, and S. Buldyrev, culminated in a joint publication \cite{Gontis2016PhysA}. Together, we investigated the intervals between volatility returns at various time scales, coining the term "volatility return intervals." Our findings revealed that the time intervals between significant financial fluctuations conform to a power-law probability density function (PDF) with $p(\tau) \sim \tau^{-3/2}$ \cite{Gontis2016PhysA}. Intriguingly, this distribution aligns with patterns observed in our models and numerous other one-dimensional Markov processes \cite{Redner2001Cambridge}. Conversely, a long-range memory process would exhibit a distinct distribution, such as $p(\tau) \sim \tau^{2-H}$, a well-known characteristic of fractional Brownian motion (FBM) \cite{Ding1995fbm}.

A significant portion of our research focuses on modeling long-range memory phenomena and power-law statistics within complex systems. Notably, our approach diverges from traditional methodologies, as we employ Markovian models in place of non-Markovian alternatives. Despite this departure, we achieve similar statistical results by harnessing various nonlinear dependencies embedded within our models. In the context of SDEs, non-linearity induces non-stationarity in the increments of the stochastic process, consequently leading to the emergence of spurious long-range memory \cite{Bassler2006PhysA,McCauley2007PhysA}. While our models differ from those representing true long-range memory, a crucial question emerges: Do our models accurately capture the observed memory in financial markets and potentially other socio-economic complex systems?

In our recent works, documented in the papers \cite{Gontis2020JStat,Gontis2022CNSNS}, we hint at three critical components required to answer this pivotal question.

The first component involves the development of a statistical test capable of distinguishing between spurious and genuine long-range memory. Our focus has centered on the Burst Duration Analysis (BDA) method \cite{Gontis2012ACS,Gontis2017Entropy,Kononovicius2019BDJStat,Gontis2020PhysA}, demonstrating competitive performance relative to alternative approaches. The core tenet of this method posits that, for any one-dimensional Markovian random walk, the probability density function (PDF) of first-passage times should exhibit a power-law behavior with an exponent of $-3/2$, at least for specific duration. Deviations from this expected behavior may indicate the presence of authentic long-range memory. Nevertheless, it is worth noting that the effectiveness of this method may wane when dealing with non-one-dimensional stochastic processes, warranting further exploration, as exemplified in multidimensional scenarios \cite{Ruseckas2016JStatMech}.

The second component necessitates the selection of models that display both spurious and true long-range memory. Our prior research has introduced a diverse array of models reflecting spurious long-range memory. Consequently, our next steps involve formulating equivalent alternative models and exploring the characteristics of existing long-range memory models. In our publication \cite{Gontis2022CNSNS}, we delved into the estimation of long-range memory within fractional L\`{e}vy stable motion, instantiated through an ARFIMA(0,d,0) discrete process. This approach represents a generalization of fractional Brownian motion. However, additional models warrant consideration \cite{Kononovicius2022}. For instance, the multiplicative point process could be enhanced by replacing uncorrelated Gaussian noise with fractional Gaussian noise. Other correlation structures or variations in pulse duration could further extend the framework \cite{Ruseckas2003LFZ}. Alternative options encompass continuous-time random walk and complex contagion frameworks.

The third component of our investigation hinges on the availability of diverse data originating from socio-economic complex systems. While our earlier work relied heavily on high-frequency absolute return and trading activity time series, our most recent endeavors have transitioned toward the utilization of order book data sourced from LOBSTER \cite{Huang2011Lobster}. The richness of order book data invites a broader approach, aligning with the framework of Fractionally Integrated Linear Stable Motion (FLSM) or ARFIMA, capable of encompassing a broad class of anomalous diffusion processes. The copious data available in social systems, particularly within financial markets, necessitates thorough scrutiny to discern and validate fractional dynamics and long-range memory phenomena. Our preliminary findings, documented in \cite{Gontis2020JStat,Gontis2022CNSNS}, cast doubt on the straightforward interpretation of long-range memory within order flow data in financial markets. To achieve clarity in this regard, it is imperative to employ prudent estimators grounded in FLSM and ARFIMA assumptions \cite{Weron2005PRE,Magdziarz2013JPhysA,Burnecki2014JStatMech}. Only after comprehensive analysis from this perspective \cite{Gontis2020JStat,Gontis2022CNSNS} can we definitively ascertain whether the observed social system indeed exhibits genuine long-range memory or whether the observed power-law statistical properties are the outcome of potent nonlinear effects.

Our contributions to the development of Econophysics underscore the potential for enhanced understanding of social system behavior through empirical data analysis, blending stochastic (macroscopic) and agent-based (microscopic) modeling approaches. Statistical physics serves as a potent instrument, facilitating the pursuit of a coherent description at both micro and macro levels for a multitude of natural and social systems. 


\begin{thebibliography}{10}
\expandafter\ifx\csname url\endcsname\relax
  \def\url#1{\texttt{#1}}\fi
\expandafter\ifx\csname urlprefix\endcsname\relax\def\urlprefix{URL }\fi
\expandafter\ifx\csname href\endcsname\relax
  \def\href#1#2{#2} \def\path#1{#1}\fi

\bibitem{Mantegna2005QF}
R.~N. Mantegna, Presentation of the english translation of ettore majoranas
  paper: The value of statistical laws in physics and social sciences,
  Quantitative Finance 5~(2) (2005) 133--140.
\newblock \href {https://doi.org/10.1080/14697680500148174}
  {\path{doi:10.1080/14697680500148174}}.

\bibitem{Chakrabarti2006Wiley}
B.~K. Chakrabarti, A.~Chakraborti, A.~Chatterjee (Eds.), Econophysics and
  Sociophysics, Wiley, 2006.
\newblock \href {https://doi.org/10.1002/9783527610006}
  {\path{doi:10.1002/9783527610006}}.

\bibitem{Gontis2001LietFizRink}
V.~Gontis, Modelling share volume traded in financial markets, Lithuanian
  Journal of Physics 41 (2001) 551--555.

\bibitem{Gontis2002MC}
V.~Gontis, Multiplicative stochastic model of the time interval between trades
  in financial markets, Nonlinear Analysis: Modelling and Control 7 (2002)
  43--54.
\newblock \href {http://arxiv.org/abs/cond-mat/0211317}
  {\path{arXiv:cond-mat/0211317}}.

\bibitem{Kaulakys1998PRE}
B.~Kaulakys, T.~Meskauskas, Modeling $1/f$ noise, Physical Review E 58 (1998)
  7013--7019.
\newblock \href {https://doi.org/10.1103/PhysRevE.58.7013}
  {\path{doi:10.1103/PhysRevE.58.7013}}.

\bibitem{Kaulakys1999PLA}
B.~Kaulakys, Autoregressive model of 1/f noise, Physics Letters A 257 (1999)
  37--42.
\newblock \href {https://doi.org/10.1016/S0375-9601(99)00284-4}
  {\path{doi:10.1016/S0375-9601(99)00284-4}}.

\bibitem{Kaulakys2005PhysRevE}
B.~Kaulakys, V.~Gontis, M.~Alaburda, Point process model of 1/f noise vs a sum
  of {L}orentzians, Physical Review E 71 (2005) 1--11.
\newblock \href {https://doi.org/10.1103/PhysRevE.71.051105}
  {\path{doi:10.1103/PhysRevE.71.051105}}.

\bibitem{Ruseckas2010PhysRevE}
J.~Ruseckas, B.~Kaulakys, 1/f noise from nonlinear stochastic differential
  equations, Physical Review E 81 (2010) 031105.
\newblock \href {https://doi.org/10.1103/physreve.81.031105}
  {\path{doi:10.1103/physreve.81.031105}}.

\bibitem{Gontis2017Entropy}
V.~Gontis, A.~Kononovicius, Spurious memory in non-equilibrium stochastic
  models of imitative behavior, Entropy 19 (2017) 387.
\newblock \href {https://doi.org/10.3390/e19080387}
  {\path{doi:10.3390/e19080387}}.

\bibitem{Gontis2018PhysA}
V.~Gontis, A.~Kononovicius, The consentaneous model of the financial markets
  exhibiting spurious nature of long-range memory, Physica {A} 505 (2018)
  1075--1083.
\newblock \href {https://doi.org/10.1016/j.physa.2018.04.053}
  {\path{doi:10.1016/j.physa.2018.04.053}}.

\bibitem{Gontis2010Intech}
V.~Gontis, J.~Ruseckas, A.~Kononovicius, A non-linear stochastic model of
  return in financial markets, in: C.~Myers (Ed.), Stochastic Control, InTech,
  2010, pp. 559--580.
\newblock \href {https://doi.org/10.5772/9748} {\path{doi:10.5772/9748}}.

\bibitem{Gontis2010PhysA}
V.~Gontis, J.~Ruseckas, A.~Kononovicius, A long-range memory stochastic model
  of the return in financial markets, Physica A 389 (2010) 100--106.
\newblock \href {https://doi.org/10.1016/j.physa.2009.09.011}
  {\path{doi:10.1016/j.physa.2009.09.011}}.

\bibitem{Gontis2011JDySES}
V.~Gontis, A.~Kononovicius, Nonlinear stochastic model of return matching to
  the data of {N}ew {Y}ork and {V}ilnius stock exchanges, Dynamics of
  Socio-Economic Systems 2 (2011) 101--109.
\newblock \href {http://arxiv.org/abs/1003.5356} {\path{arXiv:1003.5356}}.

\bibitem{Ruseckas2012ACS}
J.~Ruseckas, V.~Gontis, B.~Kaulakys, Nonextensive statistical mechanics
  distributions and dynamics of financial observables from the nonlinear
  stochastic differential equations, Advances in Complex Systems 15 (2012)
  1250073.
\newblock \href {https://doi.org/10.1142/S0219525912500737}
  {\path{doi:10.1142/S0219525912500737}}.

\bibitem{Ruseckas2011EPL}
J.~Ruseckas, B.~Kaulakys, V.~Gontis, Herding model and 1/f noise, EPL 96 (2011)
  60007.
\newblock \href {https://doi.org/10.1209/0295-5075/96/60007}
  {\path{doi:10.1209/0295-5075/96/60007}}.

\bibitem{Kononovicius2012PhysA}
A.~Kononovicius, V.~Gontis, Agent based reasoning for the non-linear stochastic
  models of long-range memory, Physica A 391 (2012) 1309--1314.
\newblock \href {https://doi.org/10.1016/j.physa.2011.08.061}
  {\path{doi:10.1016/j.physa.2011.08.061}}.

\bibitem{Gontis2014PlosOne}
V.~Gontis, A.~Kononovicius, Consentaneous agent-based and stochastic model of
  the financial markets, PLoS ONE 9 (2014) e102201.
\newblock \href {https://doi.org/10.1371/journal.pone.0102201}
  {\path{doi:10.1371/journal.pone.0102201}}.

\bibitem{Kononovicius2019OB}
A.~Kononovicius, J.~Ruseckas, Order book model with herding behavior exhibiting
  long-range memory, {Physica A} 525 (2019) 171--191.
\newblock \href {https://doi.org/10.1016/j.physa.2019.03.059}
  {\path{doi:10.1016/j.physa.2019.03.059}}.

\bibitem{Kononovicius2017Complexity}
A.~Kononovicius, Empirical analysis and agent-based modeling of {L}ithuanian
  parliamentary elections, Complexity 2017 (2017) 7354642.
\newblock \href {https://doi.org/10.1155/2017/7354642}
  {\path{doi:10.1155/2017/7354642}}.

\bibitem{Kononovicius2019CompJStat}
A.~Kononovicius, Compartmental voter model, Journal of Statistical Mechanics
  2019 (2019) 103402.
\newblock \href {https://doi.org/10.1088/1742-5468/ab409b}
  {\path{doi:10.1088/1742-5468/ab409b}}.

\bibitem{Kononovicius2020JStatMech}
A.~Kononovicius, Noisy voter model for the anomalous diffusion of parliamentary
  presence, Journal of Statistical Mechanics 2020 (2020) 063405.
\newblock \href {https://doi.org/10.1088/1742-5468/ab8c39}
  {\path{doi:10.1088/1742-5468/ab8c39}}.

\bibitem{Kononovicius2021CSF}
A.~Kononovicius, Supportive interactions in the noisy voter model, Chaos,
  Solitons {\&} Fractals 143 (2021) 110627.
\newblock \href {https://doi.org/10.1016/j.chaos.2020.110627}
  {\path{doi:10.1016/j.chaos.2020.110627}}.

\bibitem{Gontis2016PhysA}
V.~Gontis, S.~Havlin, A.~Kononovicius, B.~Podobnik, H.~E. Stanley, Stochastic
  model of financial markets reproducing scaling and memory in volatility
  return intervals, Physica A 462 (2016) 1091--1102.
\newblock \href {https://doi.org/10.1016/j.physa.2016.06.143}
  {\path{doi:10.1016/j.physa.2016.06.143}}.

\bibitem{Redner2001Cambridge}
S.~Redner, A guide to first-passage processes, Cambridge University Press,
  2001.

\bibitem{Ding1995fbm}
M.~Ding, W.~Yang, Distribution of the first return time in fractional
  {B}rownian motion and its application to the study of on-off intermittency,
  Physical Review E 52 (1995) 207.
\newblock \href {https://doi.org/10.1103/PhysRevE.52.207}
  {\path{doi:10.1103/PhysRevE.52.207}}.

\bibitem{Bassler2006PhysA}
K.~E. Bassler, G.~H. Gunaratne, J.~L. McCauley, Markov processes, {H}urst
  exponents, and nonlinear diffusion equations: {W}ith application to finance,
  Physica A 369 (2006) 343--353.
\newblock \href {https://doi.org/10.1016/j.physa.2006.01.081}
  {\path{doi:10.1016/j.physa.2006.01.081}}.

\bibitem{McCauley2007PhysA}
J.~L. McCauley, G.~H. Gunaratne, K.~E. Bassler, Hurst exponents, {M}arkov
  processes, and fractional {B}rownian motion, Physica A 379 (2007) 1--9.
\newblock \href {https://doi.org/10.1016/j.physa.2006.12.028}
  {\path{doi:10.1016/j.physa.2006.12.028}}.

\bibitem{Gontis2020JStat}
V.~Gontis, Long-range memory test by the burst and inter-burst duration
  distribution, Journal of Statistical Mechanics 2020 (2020) 093406.
\newblock \href {https://doi.org/10.1088/1742-5468/abb4db}
  {\path{doi:10.1088/1742-5468/abb4db}}.

\bibitem{Gontis2022CNSNS}
V.~Gontis, Order flow in the financial markets from the perspective of the
  fractional l{\'{e}}vy stable motion, Communications in Nonlinear Science and
  Numerical Simulation 105 (2022) 106087.
\newblock \href {https://doi.org/10.1016/j.cnsns.2021.106087}
  {\path{doi:10.1016/j.cnsns.2021.106087}}.

\bibitem{Gontis2012ACS}
V.~Gontis, A.~Kononovicius, S.~Reimann, The class of nonlinear stochastic
  models as a background for the bursty behavior in financial markets, Advances
  in Complex Systems 15 (2012) 1250071.
\newblock \href {https://doi.org/10.1142/S0219525912500713}
  {\path{doi:10.1142/S0219525912500713}}.

\bibitem{Kononovicius2019BDJStat}
A.~Kononovicius, V.~Gontis, Approximation of the first passage time
  distribution for the birth-death processes, Journal of Statistical Mechanics
  2019 (2019) 073402.
\newblock \href {https://doi.org/10.1088/1742-5468/ab2709}
  {\path{doi:10.1088/1742-5468/ab2709}}.

\bibitem{Gontis2020PhysA}
V.~Gontis, A.~Kononovicius, Bessel-like birth-death process, Physica A 540
  (2020) 123119.
\newblock \href {https://doi.org/10.1016/j.physa.2019.123119}
  {\path{doi:10.1016/j.physa.2019.123119}}.

\bibitem{Ruseckas2016JStatMech}
J.~Ruseckas, R.~Kazakevicius, B.~Kaulakys, Coupled nonlinear stochastic
  differential equations generating arbitrary distributed observable with 1/f
  noise, Journal of Statistical Mechanics 2016 (2016) 043209.
\newblock \href {https://doi.org/10.1088/1742-5468/2016/04/043209}
  {\path{doi:10.1088/1742-5468/2016/04/043209}}.

\bibitem{Kononovicius2022}
A.~Kononovicius, R.~Kazakevičius, B.~Kaulakys,
 Resemblance of the power-law scaling behavior of a non-markovian and nonlinear point
  processes, Chaos, Solitons \& Fractals 162 (2022) 112508.
\newblock \href {https://doi.org/doi.org/10.1016/j.chaos.2022.112508}
  {\path{doi:doi.org/10.1016/j.chaos.2022.112508}}.
\newline\urlprefix\url{https://www.sciencedirect.com/science/article/pii/S0960077922007111}

\bibitem{Ruseckas2003LFZ}
J.~Ruseckas, B.~Kaulakys, M.~Alaburda, Modelling of $1/f$ noise by sequences of
  stochastic pulses of different duration, Lithuanian Journal of Physics 43
  (2003) 223--228.
\newblock \href {http://arxiv.org/abs/0812.4674} {\path{arXiv:0812.4674}}.

\bibitem{Huang2011Lobster}
R.~Huang, T.~Polak, \href{https://lobsterdata.com/LobsterReport.pdf}{{LOBSTER}:
  The limit order book reconstructor}, Tech. rep., Humboldt Universitat zu
  Berlin, discussion paper School of Business and Economics (2011).
\newline\urlprefix\url{https://lobsterdata.com/LobsterReport.pdf}

\bibitem{Weron2005PRE}
A.~Weron, K.~Burnecki, Complete description of all self-similar models driven
  by {L}evy stable noise, Physical Review E 71 (2005) 016113.
\newblock \href {https://doi.org/10.1103/PhysRevE.71.016113}
  {\path{doi:10.1103/PhysRevE.71.016113}}.

\bibitem{Magdziarz2013JPhysA}
M.~Magdziarz, J.~K. Slezak, J.~Wojcik, Estimation and testing of the {H}urst
  parameter using p-variation, Journal of Physics A: Mathematical and
  Theoretical 46 (2013) 325003.
\newblock \href {https://doi.org/10.1088/1751-8113/46/32/325003}
  {\path{doi:10.1088/1751-8113/46/32/325003}}.

\bibitem{Burnecki2014JStatMech}
K.~Burnecki, A.~Weron, Algorithms for testing of fractional dynamics: {A}
  practical guide to {ARFIMA} modelling, Journal of Statistical Mechanics 2014
  (2014) P10036.
\newblock \href {https://doi.org/10.1088/1742-5468/2014/10/p10036}
  {\path{doi:10.1088/1742-5468/2014/10/p10036}}.

\end{thebibliography}

\end{document}